# Fault-Tolerant Control of a 2 DOF Helicopter (TRMS System) Based on H∞

Abderrahmen BOUGUERRA[#1], Djamel SAIGAA[#2], Kamel KARA [*1]

Samir ZEGHLACHE[#3], Keltoum LOUKAL[#4]

[#1, 3, 4] *LASS Laboratory, Department of Electrical engineering, University of M'sila,*

[#2] *LASS Laboratory, Department of Electronics, University of M'sila,*

*BP 166 Ichbilia 28000 Algeria*

[1]rah_bou@yahoo.fr

[2]saigaa_dj@yahoo.fr

[3]zeghlache_samir@yahoo.fr

[4]muohtlek@yahoo.fr

[*] *Department of Electronics, University of Blida,*

*Route De Soumaa BP 270 Blida Algeria*

[1]kara_k_dz@yahoo.fr

*Abstract*— **In this paper, a Fault-Tolerant control of 2 DOF Helicopter (TRMS System) Based on H∞ is presented. In particular, the introductory part of the paper presents a Fault-Tolerant Control (FTC), the first part of this paper presents a description of the mathematical model of TRMS, and the last part of the paper presented and a polytypic Unknown Input Observer (UIO) is synthesized using equalities and LMIs. This UIO is used to observe the faults and then compensate them, in this part the shown how to design a fault-tolerant control strategy for this particular class of non-linear systems.**

*Keywords*— **Helicopter model; H∞ control; UIO; state feedback control; MM; FTC.**

## I. Introduction

Fault-Tolerant Control (FTC) is a relatively new idea that makes possible to develop a control feedback that allows keeping the required system performance in the case of faults [1]. The control strategy can be perceived fault tolerant when there is an adaptation mechanism that changes the control law in the case of faults. Another solution is to use hardware redundancy in sensors and/or actuators. In general, FTC systems are classified into two distinct classes [2]: passive and active. In passive FTC [3] [4], controllers are designed to be robust against a set of presumed faults, therefore there is no need for fault detection. In the contrast to passive ones, active FTC schemes, react to system components faults actively by reconfiguring control actions, and by doing so the system stability and acceptable performance is maintained.

Due to the complicated nonlinearity and the high coupling effect between two propellers, the control problem of the (TRMS) has been considered as a challenging research topic [5]. Moreover, the control of the TRMS has gained a lot of attention because the dynamics of the TRMS and a helicopter are similar in certain aspects [6], [7]. A multivariable nonlinear H∞ controller is designed in [8] for the angle control of the TRMS. The remainder of this paper is organized as follows. The model of the TRMS is described in Section II. The FTC strategy is designed in Section III. Section IV presents the simulation results to demonstrate the effectiveness of the FTC Controller. Concluding remarks are provided in Section VI.

## II. Model Description of the TRMS

Similar to most flight vehicles, the helicopter consists of several elastic parts such as rotor, engine and control surfaces. The nonlinear aerodynamic forces and gravity act on the vehicle, and flexible structures increase complexity and make a realistic analysis difficult. For control purpose, it is necessary to find a representative model that shows the same dynamic characteristics as the real aircraft [9]. The behaviour of a nonlinear TRMS, (shown in Fig.1), in certain aspects resembles that of a helicopter. It can be well perceived as a static test rig for an air vehicle with formidable control challenges.

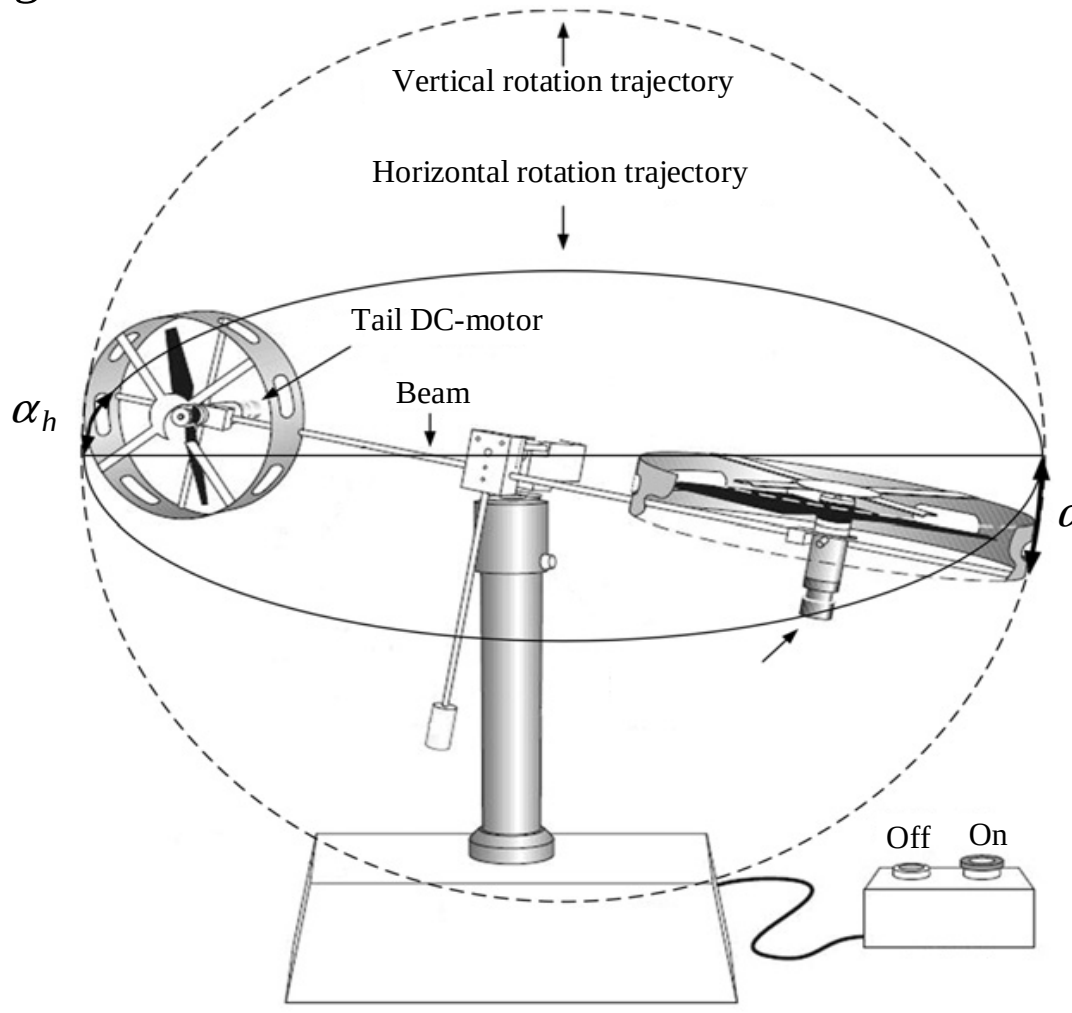

Fig.1 The twin rotor multi-input multi-output system (TRMS) [10]

This TRMS consists of a beam pivoted on its base in such a way that it can rotate freely in both its horizontal and vertical planes. There are two rotors (the main and tail rotors), driven by DC motors, at each end of the beam. If necessary, either or both axes of rotation can be locked by means of two locking screws provided for physically restricting the horizontal or vertical plane rotation. Thus, the system permits both 1 and 2 degree-of-freedom (DOF) experiments.

The two rotors are controlled by variable speed electric motors enabling the helicopter to rotate in a vertical and horizontal plane (pitch and yaw). The mathematical model of the TRMS is developed under following assumptions.

- The dynamics of the propeller subsystem can be described by first-order differential equations.
- The friction in the system is of the viscous type.
- The propeller – air subsystem could be described in accordance with the postulates of the flow theory.

The mechanical system of TRMS is simplified using a four point-mass system shown in Fig. 2.

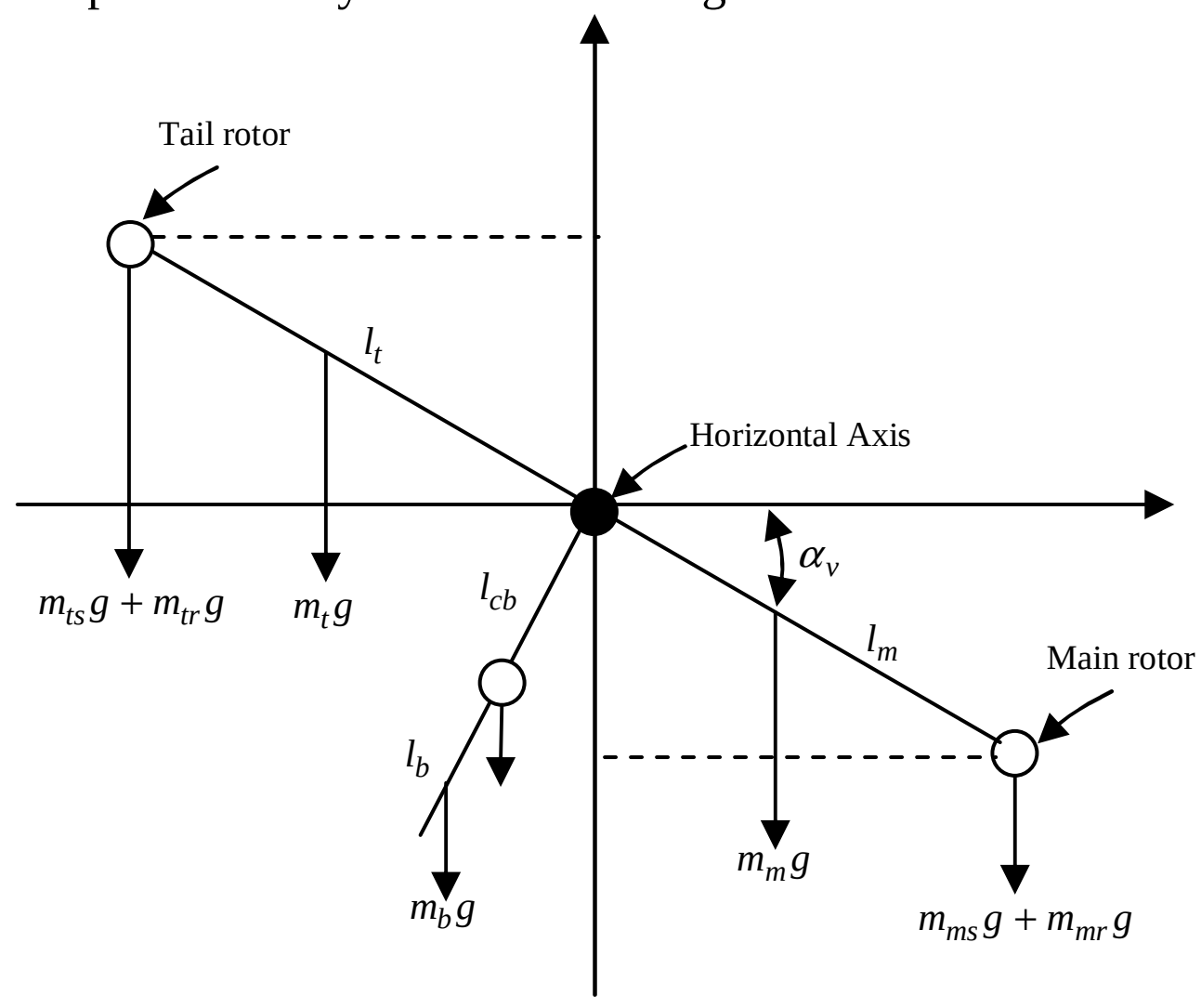

Fig.2 Simplified four point-mass systems

The parameters in the simplified four point-mass system are $M_{v1}$ is the return torque corresponding to the force of gravity, $M_{v2}$ is the moment of a aerodynamic force, $M_{v3}$ is the moment of a centrifugal forces, $M_{v4}$ is a Moment of friction, $m_{mr}$ is the mass of the DC motor within the main rotor, $m_m$ is the mass of the main part of the beam, $m_{tr}$ is the mass of the DC motor within tail rotor, $m_t$ is the mass of the tail part of the beam, $m_{cb}$ is the mass of the counter weight, $m_b$ is the mass of the counter-weight beam, $m_{ms}$ is the mass of the main shield, $m_{ts}$ is the mass of the tail shield, $l_m$ is the length of the main part of the beam, $l_t$ is the length of the tail part of the beam, $l_b$ is the length of the counter-weight beam, $l_{cb}$ is the distance between the counter-weight and joint, and g is the gravitational acceleration.

The driving torqueses are produced by the propellers, and the rotation can be described in principle as the motion of a pendulum. We can write the equations describing this motion as follows.

*A. The main rotor model*

$$M_{v1} = g\left\{ \begin{array}{l} \left[\left(\frac{m_t}{2} + m_{tr} + m_{ts}\right) l_t - \left(\frac{m_m}{2} + m_{mr} + m_{ms}\right) l_m\right] \cos\alpha_v - \\ \left(\frac{m_b}{2} l_b + m_{cb}\ \ l_{cb}\right) \sin\alpha_v \end{array} \right\} \quad (1)$$

$$M_{v1} = g\ \{[A - B]\cos\alpha_v - C\sin\alpha_v\} \quad (2)$$

With:

$$\begin{cases} A = \left(\frac{m_t}{2} + m_{tr} + m_{ts}\right) l_t \\ B = \left(\frac{m_m}{2} + m_{mr} + m_{ms}\right) l_m \\ C = \left(\frac{m_b}{2} l_b + m_{cb}\ \ l_{cb}\right) \end{cases} \quad (3)$$

$$M_{v2} = l_m S_f F_v(\omega_m) \quad (4)$$

The angular velocity $\omega_m$ of main propeller is a nonlinear function of a rotation angle of the DC motor describing by:

$$\begin{array}{l} \omega_m(u_{vv}) = 90.90\,u_{vv}^6 + 599.73\,u_{vv}^5 - 129.26\,u_{vv}^4 - \\ \quad 1238.64\,u_{vv}^3 + 63.45\,u_{vv}^2 + 1238.41\,u_{vv} \end{array} \quad (5)$$

Also, the propulsive force $F_v$ moving the joined beam in the vertical direction is describing by a nonlinear function of the angular velocity $\omega_m$.

$$\begin{array}{l} F_v(\omega_m) = -3.48\times10^{-12}\omega_m^5 + 1.09\times10^{-9}\omega_m^4 + 4.123\times10^{-6}\omega_m^3 - \\ \quad 1.632\times10^{-4}\omega_m^2 + 9.544\times10^{-2}\omega_m \end{array} \quad (6)$$

The model of the motor-propeller dynamics is obtained by substituting the nonlinear system by a serial connection of a linear dynamics system. This can be expressed as:

$$\frac{du_{vv}}{dt} = \frac{1}{T_{mr}}\left(-u_{vv} + u_v\right) \quad (7)$$

$u_v$ is the input voltage of the DC motor, $T_{mr}$ is the time constant of the main rotor and $K_{mr}$ is the static gain DC motor.

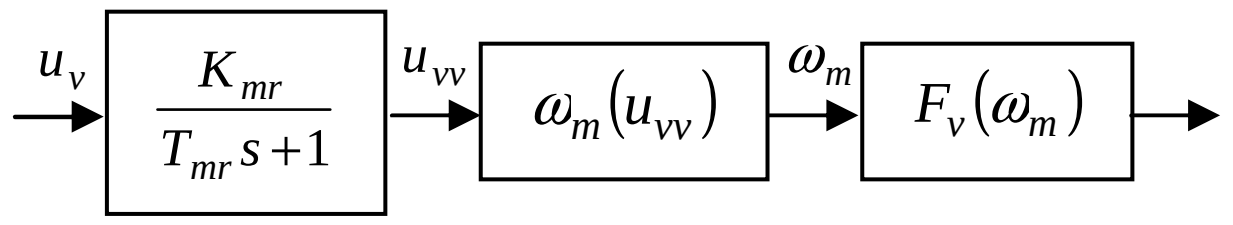

Fig 3 The relationship between the input voltage and the propulsive force for the main rotor

$$M_{v3} = -\Omega_h^2 \left\{ \begin{array}{l} \left(\frac{m_t}{4} + m_{tr} + m_{ts}\right) l_t^2 + \left(\frac{m_t}{4} + m_{tr} + m_{ts}\right) l_m^2 - \\ \left(\frac{m_b}{4} l_b^2 + m_{cb} l_{cb}^2\right) \end{array} \right\} \sin\alpha_v \cos\alpha_v \quad (8)$$

$$M_{v3} = -\Omega_h^2 (H) \sin \alpha_v \cos \alpha_v \quad (9)$$

With:

$$H = \left(\frac{m_t}{4} + m_{tr} + m_{ts}\right) l_t^2 + \left(\frac{m_t}{4} + m_{tr} + m_{ts}\right) l_m^2 - \left(\frac{m_b}{4} l_b^2 + m_{cb} l_{cb}^2\right) \quad (10)$$

$$\Omega_h = \frac{d\alpha_h}{dt} \quad (11)$$

$$M_{v4} = -\Omega_v K_v \quad (12)$$

$$\Omega_v = \frac{d\alpha_v}{dt} \quad (13)$$

$$\frac{dS_v}{dt} = \frac{1}{J_v} \sum_{i=1}^{4} M_{vi} \quad (14)$$

$$\frac{dS_v}{dt} = \frac{1}{J_v} \{ l_m S_f F_v(\omega_m) - K_v \Omega_v + g((A-B)\cos \alpha_v - C \sin \alpha_v) - \Omega_h^2 (H) \sin \alpha_v \cos \alpha_v \} \quad (15)$$

$$\frac{d\alpha_v}{dt} = \Omega_v = S_v + \frac{J_{tr} \omega_t}{J_v} \quad (16)$$

Where $\omega_t$ is the angular velocity of tail propeller, $S_v$ the angular momentum in the vertical plane of the beam, $J_v$ the sum of inertia moments in the horizontal plane, $J_{tr}$ the moment of inertia in DC motor tail propeller subsystem, $K_v$ the Friction constant, and $S_f$ the balance scale.

*B. The tail rotor model*

Similarly, we can describe the motion of the beam in the horizontal plane (around the vertical axis) as shown in Fig.4. The driving torqueses are produces by the rotors and that the moment of inertia depends on the pitch angle of the beam.

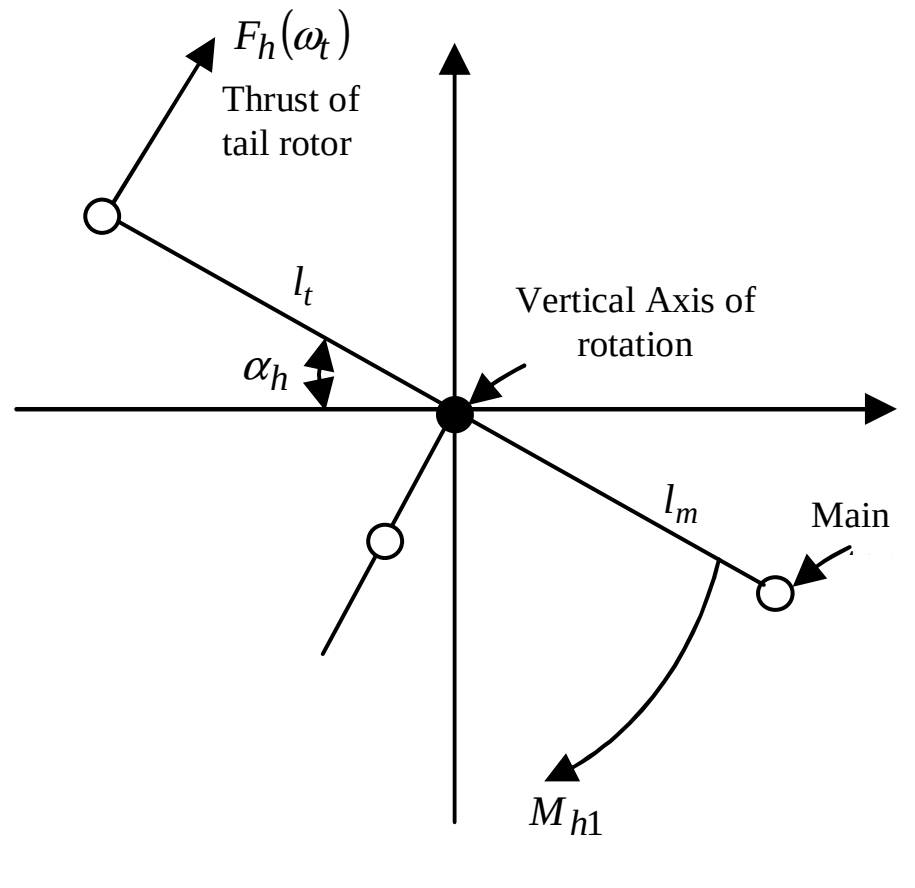

Fig 4 Torques around the vertical axis

The parameters in the torques around vertical axis are $M_{h1}$ is the moment of an aerodynamic force, $M_{h2}$ is a Moment of friction.

$$M_{h1} = l_t S_f F_h(\omega_t) \cos \alpha_v \quad (17)$$

The angular velocity $\omega_t$ of tail propeller is a nonlinear function of a rotation angle of the DC motor describing by:

$$\omega_t(u_{hh}) = 2020\, u_{hh}^5 + 194.69\, u_{hh}^4 - 4283.15\, u_{vv}^3 - 262.87\, u_{hh}^2 + 3796.83\, u_{hh} \quad (18)$$

Also, the propulsive force $F_h$ moving the joined beam in the Horizontal direction is describing by a nonlinear function of the angular velocity $\omega_t$

$$F_h(\omega_t) = -3\times10^{-14} \omega_t^5 + 1.595\times10^{-11} \omega_t^4 + 2.511\times10^{-7} \omega_t^3 - 1.808\times10^{-4} \omega_t^2 + 0.8080\, \omega_t \quad (19)$$

The model of the motor-propeller dynamics is obtained by substituting the nonlinear system by a serial connection of a linear dynamics system. This can be expressed as:

$$\frac{du_{hh}}{dt} = \frac{1}{T_{tr}} (-u_{hh} + u_h) \quad (20)$$

$u_h$ is the input voltage of the DC motor, $T_{tr}$ is the time constant of the tail rotor and $K_{tr}$ is the static gain DC motor .

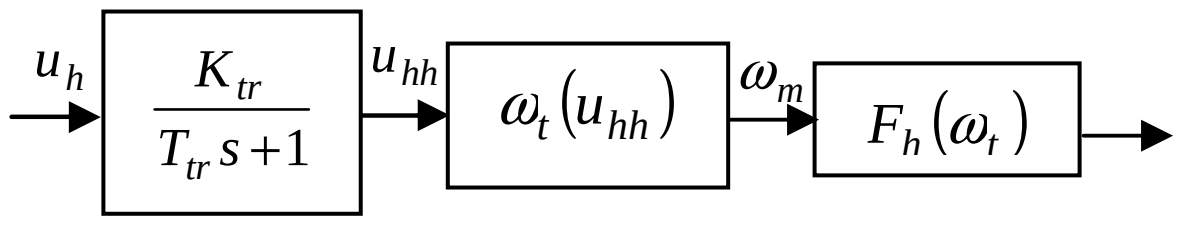

Fig 5 The relationship between the input voltage and the propulsive force for the tail rotor

$$M_{h2} = -\Omega_h K_h \quad (21)$$

$$\frac{dS_h}{dt} = \frac{1}{J_h(\alpha_v)} \sum_{i=1}^{4} M_{hi} \quad (22)$$

$$J_h(\alpha_v) = D \cos^2 \alpha_v + E \sin^2 \alpha_v + F \quad (23)$$

$$\frac{dS_h}{dt} = \frac{l_t S_f F_h(\omega_t) \cos \alpha_v - \Omega_h K_h}{J_h(\alpha_v)} \quad (24)$$

$$\frac{dS_h}{dt} = \frac{l_t S_f F_h(\omega_t) \cos \alpha_v - \Omega_h K_h}{D \cos^2 \alpha_v + E \sin^2 \alpha_v + F} \quad (25)$$

$$\frac{d\alpha_h}{dt} = \Omega_h = S_h + \frac{J_{mr}\, \omega_m \cos \alpha_v}{D \cos^2 \alpha_v + E \sin^2 \alpha_v + F} \quad (26)$$

Where $S_h$ the angular momentum in the horizontal plane of the beam, $J_h$ the sum of inertia moments in the vertical plane, $J_{mr}$ the moment of inertia in DC motor main propeller subsystem, $K_h$ the Friction constant, and $S_f$ the balance scale.

The model in the state-space is:
$\dot{X} = f(x) + g(X, U)$ and $X = [x_1, \ldots, x_6]^T$ is the state vector of the system such as:

$$X = [\alpha_v, S_v, u_{vv}, \alpha_h, S_h, u_{hh}] \quad (27)$$

$$U = [u_v, u_h] \quad (28)$$

$$Y = [\alpha_v, \alpha_h] \tag{29}$$

From (26), (27) and (28) we obtain the following state representation:

$$\begin{cases} \dot{x}_1 = x_2 + \dfrac{J_{tr}}{J_v}\omega_t(x_6) \\ \dot{x}_2 = \dfrac{1}{J_v}\left\{ \begin{array}{l} l_m S_f F_v(\omega_m(x_3)) - K_v\left(x_2 + \dfrac{J_{tr}}{J_v}\omega_t(x_6)\right) + g((A-B)\cos x_1 - C\sin x_1) \\ -\left(x_5 + \dfrac{J_{mr}\omega_m(x_3)\cos x_1}{D\cos^2 x_1 + E\sin^2 x_1 + F}\right)^2 (H)\sin x_1 \cos x_1 \end{array} \right\} \\ \dot{x}_3 = \dfrac{1}{T_{mr}}(-x_3 + K_{mr}u_v) \\ \dot{x}_4 = x_5 + \dfrac{J_{mr}\omega_m(x_3)\cos x_1}{D\cos^2 x_1 + E\sin^2 x_1 + F} \\ \dot{x}_5 = \dfrac{1}{D\cos^2 x_1 + E\sin^2 x_1 + F}\left\{ l_t S_f F_h(\omega_t)\cos\alpha_v - K_h\left(x_5 + \dfrac{J_{mr}\omega_m(x_3)\cos x_1}{D\cos^2 x_1 + E\sin^2 x_1 + F}\right)\right\} \\ \dot{x}_6 = \dfrac{1}{T_{tr}}(-x_6 + K_{tr}u_h) \end{cases}$$

(30)

TABLE I
THE PARAMETERS OF THE TRMS [10]

| Symbol | Definition | Value |
|---|---|---|
| A | Mechanical related constant | 0.0946875 $kgm^2$ |
| B | Mechanical related constant | 0.11046 $kgm^2$ |
| C | Mechanical related constant | 0.01986 $kgm^2$ |
| D | Mechanical related constant | 0.04988 $kgm^2$ |
| E | Mechanical related constant | 0.004745 $kgm^2$ |
| F | Mechanical related constant | 0.006230 $kgm^2$ |
| H | Mechanical related constant | 0.048210 $kgm^2$ |
| $S_f$ | Balanced scale | 0.000843318 |
| $J_v$ | Sum of inertia moments in the horizontal plane | 0.055448 $kgm^2$ |
| $J_{mr}$ | Moment of inertia in the DC-motor of main propeller | 0.000016543 $kgm^2$ |
| $J_{tr}$ | Moment of inertia in the DC-motor of main propeller | 0.0000265 $kgm^2$ |
| $l_m$ | Length of the main part of the beam | 0.24 m |
| $l_t$ | Length of the tail part of the beam | 0.25 m |
| $T_{mr}$ | Time constant of the main rotor | 1.432 sec |
| $T_{tr}$ | Time constant of the tail rotor | 0.3842 sec |
| $K_{mr}$ | Static gain of the main DC-motor | 1 |
| $K_{tr}$ | Static gain of the tail DC-motor | 1 |
| $K_v$ | Friction coefficient for the vertical axis | 0.0095 |
| $K_h$ | Friction coefficient for the horizntal axis | 0.00545371 |
| g | Gravitational acceleration | 9.81 $m/s^2$ |

## III. FTC STRATEGY

Consider a system represented by the Multi model:

$$\begin{cases} \dot{x}(t) = \sum_{i=1}^{N}\mu_i(\xi(t))(A_i x(t) + B_i u(t) + \Delta X_i) \\ y(t) = \sum_{i=1}^{N}\mu_i(\xi(t))(C_i x(t) + \Delta Y_i) \end{cases} \tag{31}$$

Where $x(t)\epsilon R^n$ is the state vector, $y(t)\epsilon R^p$ is the output vector and $u(t)\epsilon R^m$ is the input vector $\xi(t)$. Is the vector of decision variables which can depend on the state of the outputs or inputs. Considering the system of equations (31), $S$ is a matrix of sequencing models varying as follows:

$$S = \left\{ \sum_{i=1}^{N}\mu_i S_i : \mu_i \geq 0, \sum_{i=1}^{N}\mu_i = 1 \right\} \tag{32}$$

With:

$$S_i = \begin{bmatrix} A_i & B_i & \Delta X_i \\ C_i & \Delta Y_i & \end{bmatrix}, \forall i \in \{1, ..., N\} \tag{33}$$

When fault actuators (as additive), the previous system is written as follows:

$$\begin{cases} \dot{x}_f(t) = \sum_{i=1}^{N}\mu_i(\xi(t))(A_i x_f(t) + B_i u_f(t) + L_i f + \Delta X_i) \\ y_f(t) == \sum_{i=1}^{N}\mu_i(\xi(t))(C_i x_f(t) + \Delta Y_i) \end{cases} \tag{34}$$

Where $L_i$ is the distribution matrix faults [11] [12].

### *A. Fault tolerant control in multi-model*

We consider a single output matrix $C$ for different operating points, the system (32) is written as follows:

$$\begin{cases} \dot{x}(t) = \sum_{i=1}^{N}\mu_i(\xi(t))(A_i x(t) + B_i u(t) + \Delta X_i) \\ y(t) == C x(t) \end{cases} \tag{35}$$

Considering the additive representation actuators faults, the system (36) takes the following form:

$$\begin{cases} \dot{x}_f(t) = \sum_{i=1}^{N}\mu_i(\xi(t))(A_i x_f(t) + B_i u_f(t) + L_i f(t) + \Delta X_i) \\ y_f(t) = C x_f(t) \end{cases} \tag{36}$$

Where $x_f(t)\epsilon R^n$ is the state vector, $y_f(t)\epsilon R^p$ is the output vector and $u_f(t)\epsilon R^m$ is the input vector. The state matrix of $i^{th}$ local model is $A_i\epsilon R^{n\times n}$ the control matrix is $B_i\epsilon R^{n\times m}$ and the output matrix is $C\epsilon R^{p\times n}$ Distribution matrices faults are noted $L_i\epsilon R^{n\times s}$ supposed full column rank, and $f\epsilon R^s$ represents the vector of faults.

The term $\Delta X_i\epsilon R^{n\times 1}$ a vector is dependent $i^{th}$ operating point, we assume that the matrices $L_i, \forall i \in \{1, ..., N\}$ i.e. less if $f \neq 0$ then $L_i f \neq 0$, In addition, it should be noted that in a conventional manner for detectability of faults. The objective of the method is to synthesize a control law $u_f(t)$ which cancels the effect of defects on the system Closed loop converges asymptotically and state $x_f(t)$ to state $x(t)$ despite the presence of faults. The control law is synthesized as follows [11]:

$$u_f(t)=\sum_{i=1}^{N}\mu_i(\xi(t))(-S_i\hat{f}(t)+K_{1,i}(x(t)-x_f(t))+u(t)) \quad (37)$$

Where $\hat{f}(t)$ represents an estimate of the fault. The purpose of the first term of the command $S_i\hat{f}(t)$ is the estimate of default, the purpose of the term $K_{1,i}\left(x(t)-x_f(t)\right)$ is to annul the estimation error and $u(t)$ defines the nominal command. Replacing the state vectors $x(t)$ and $x_f(t)$ by their estimated $\hat{x}(t)$ and $\hat{x}_f(t)$ in equation (37) the application of this technique is equivalent to:

- Determine $\hat{f}(t)$.
- Calculate the gains $K_{1,i}, \forall i \in \{1,\dots,N\}$ such that the closed loop system is stable.

## B. *Faults Estimation*

Actuators faults represented by the vector $f(t)$ in equation (37) can be considered as unknown inputs. This allows us to use the theory of unknown input observers for UIO estimate. And state estimation in the presence of a fault condition tends to itself. Then replaced $x_f(t)$ by its estimate $\hat{x}_f(t)$ in equation (37) and we obtain:

$$\begin{cases}\dot{\hat{x}}_f(t)=\sum_{i=1}^{N}\mu_i(\xi(t))(A_i\,\hat{x}_f(t)+B_i\,u_f(t)+L_i\,\hat{f}(t)+\Delta X_i)\\ \hat{y}_f(t)=C\,\hat{x}_f(t)\end{cases} \quad (38)$$

The expression of the fault estimation error is:

$$f(t)-\hat{f}(t)=-\sum_{i=1}^{N}\mu_i(\xi(t))\,H_iC\,A_i(x_f(t)-\hat{x}_f(t)) \quad (39)$$

With:

$$H_i=\left[(C\,L_i)^T C\,L_i\right]^{-1}(C\cdots L_i)^T \quad (40)$$

## C. *Synthesis of the fault tolerant control*

The synthetic methodology of proposed FTC command was built around the following assumptions:

- The pairs $(A_i, C)$ are observable.
- The pairs $(A_i, B)$ are controllable.

The first assumption is necessary for calculating the gains of multiple observers, whereas the second allows the calculation of the command gains FTC. Thus it is possible to calculate the matrices $K_{1,i}$ et $K_{2,i}$ .

$$\dot{\bar{e}}(t)=\sum_{i=1}^{N}\mu_i(\xi(t))\begin{bmatrix}A_i-B_i\,K_{1,i} & L_iH_iCA_i\\ 0 & \bar{A}_i-K_{2,i}C\end{bmatrix}\bar{e}(t) \quad (41)$$

$$\dot{\bar{e}}(t)=\sum_{i=1}^{N}\mu_i(\xi(t))\,A_{0,i}\bar{e}(t) \quad (42)$$

After several simulations choice is focused on the following parameters:

$\zeta = 2$ et $\rho = 700$

We obtain the following matrix gain for the first three points operation:

$$\begin{cases}K_{11}=10^3\begin{pmatrix}1.2333 & 0.0423 & -0.0102 & 6.2134 & 1.8324 & 0.0107\\ 0.1113 & -0.0547 & -0.0105 & 1.0413 & 0.5248 & 0.0005\end{pmatrix}\\ K_{12}=10^4\begin{pmatrix}0.0044 & 0.2623 & 0.0109 & -1.2356 & -1.3388 & -0.1001\\ 0.0114 & -0.0237 & 0.2222 & 0.0742 & 0.1548 & 0.0004\end{pmatrix}\\ K_{13}=10^4\begin{pmatrix}0.2865 & 0.1252 & -0.1008 & 1.3628 & 0.7540 & 0.0007\\ 0.0140 & -0.1021 & 0.0004 & 0.1744 & 0.0319 & 0.9932\end{pmatrix}\end{cases} \quad (43)$$

The matrices of of multi-observer gains are given by:

$$\bar{A}_1=\begin{pmatrix}0 & 1 & 0 & 0 & 0 & 0\\ -2.0623 & -1.0123 & 0.0044 & 0 & 0 & 0\\ 0 & 0 & 0 & 0 & 0 & 0\\ 0 & 0 & 0 & 0 & 1 & 0\\ 0 & & 0 & 0 & -1.2077 & 0.1001\\ 0 & 0 & 0 & 0 & 0 & 0\end{pmatrix}, \bar{L}_1=\begin{pmatrix}0 & 0 & 0 & 0\\ 0 & 0 & 0 & 0.0002\\ 0 & 1 & 0 & 0\\ 0 & 0 & 0 & 0\\ 0 & 0.0003 & 0 & 0\\ 0 & 0 & 0 & 1\end{pmatrix}$$

$$K_{2,1}=\begin{pmatrix}'38.3682 & 0 & -1.1060 & 0\\ 527.4795 & 0.0048 & -7.7743 & 0\\ 0 & 25.4456 & 0 & 0\\ -0.1389 & 0 & 33.2744 & 0\\ -0.8171 & 0 & 52.3369 & 0.1083\\ 0 & 0 & 0 & 29.6656\end{pmatrix}, \bar{B}_1=10^{-11}\begin{pmatrix}0 & 0\\ 0 & 0\\ 0 & 0\\ 0 & 0\\ 0 & 0\\ 0 & 0.4398\end{pmatrix}, \Delta\bar{X}_1=\begin{pmatrix}0\\ -8.2922\\ 0\\ 0.4176\\ 0\end{pmatrix}$$
(44)

# IV. SIMULATION RESULTS

The proposed faults tolerant control scheme presented in this paper was tested on a model of helicopter, which is called a twin rotor MIMO system Fig. 5.

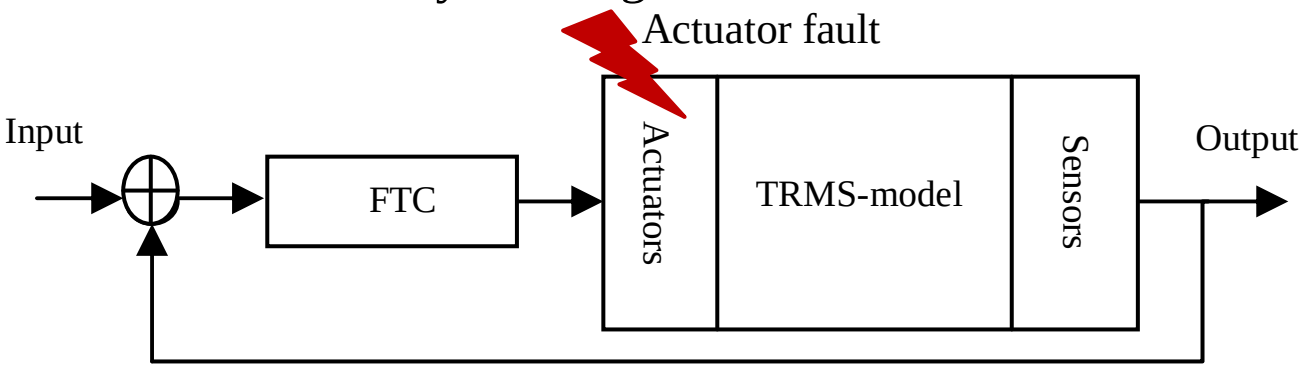

Fig. 5 faults actuators in the control loop

The values of the main mechanical parameters of this system are listed in Table I. in order to show the effectiveness of the proposed approach. Four simulations are performed. The system was simulated selecting an intermittent fault is triggered at the time 25s, and it is estimated simultaneously in the multiple observer unknown input. Fig. 8 and Fig. 9 show the fault and its estimate.

## A. *nominal control by state feedback with FTC*

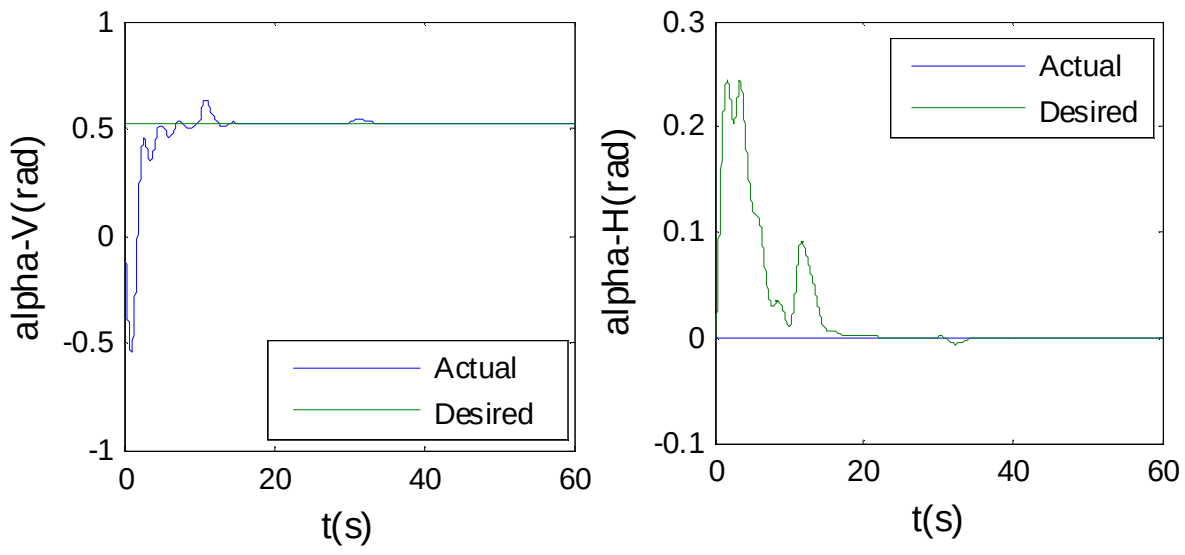

Fig. 6 the vertical and horizontal angles tracking

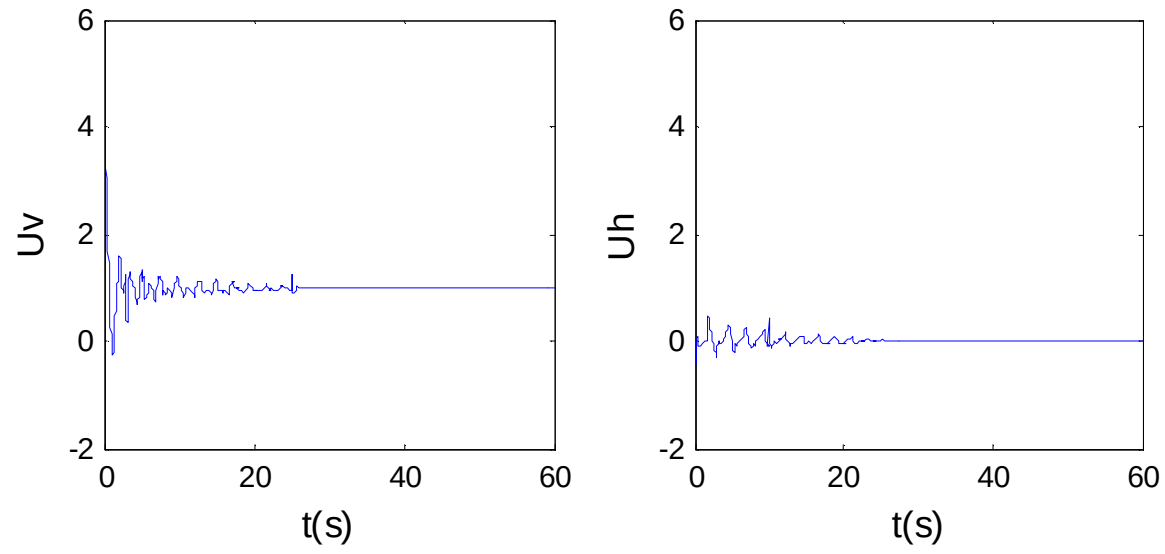

Fig. 7 system commands

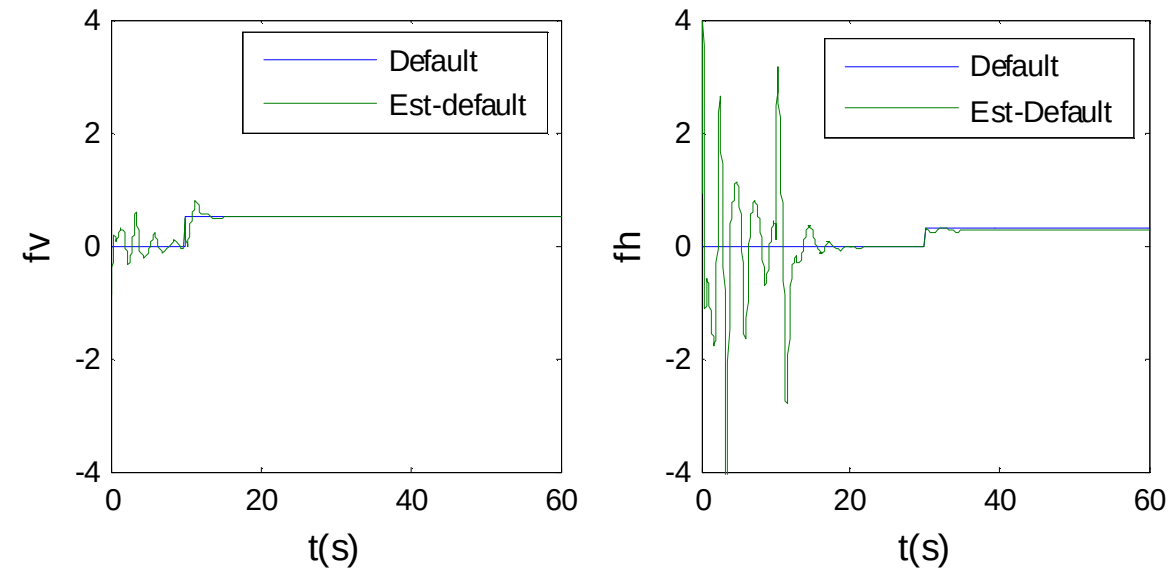

Fig. 8 Defaults: f(t) and the estimated default f̂(t)

### *B. nominal control $H_\infty$ with FTC*

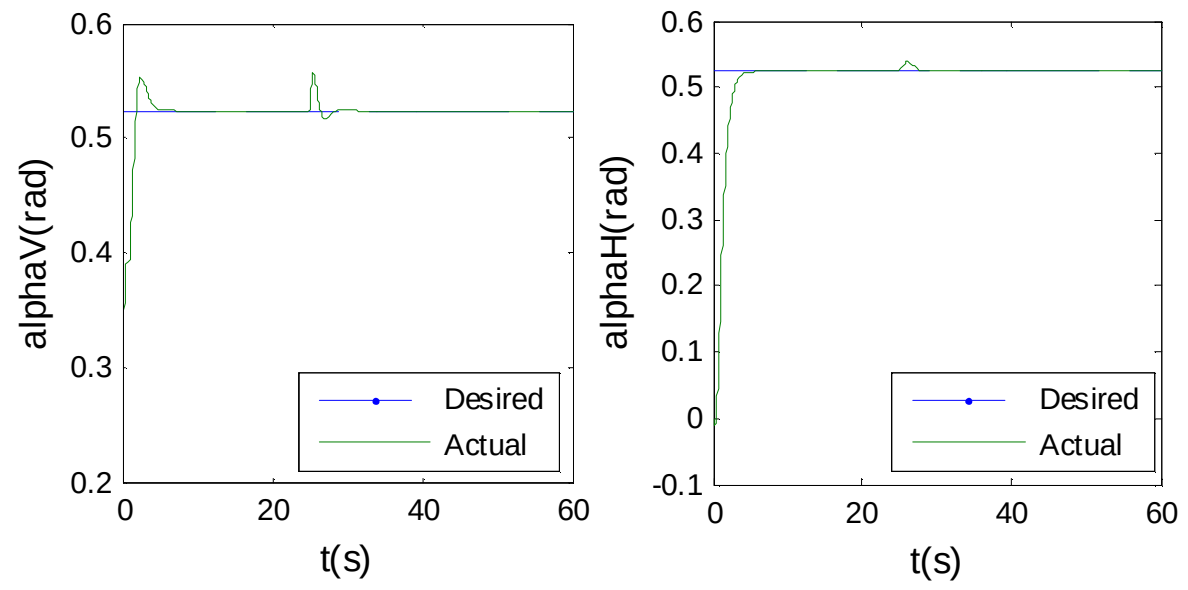

Fig. 9 the vertical and horizontal angles tracking

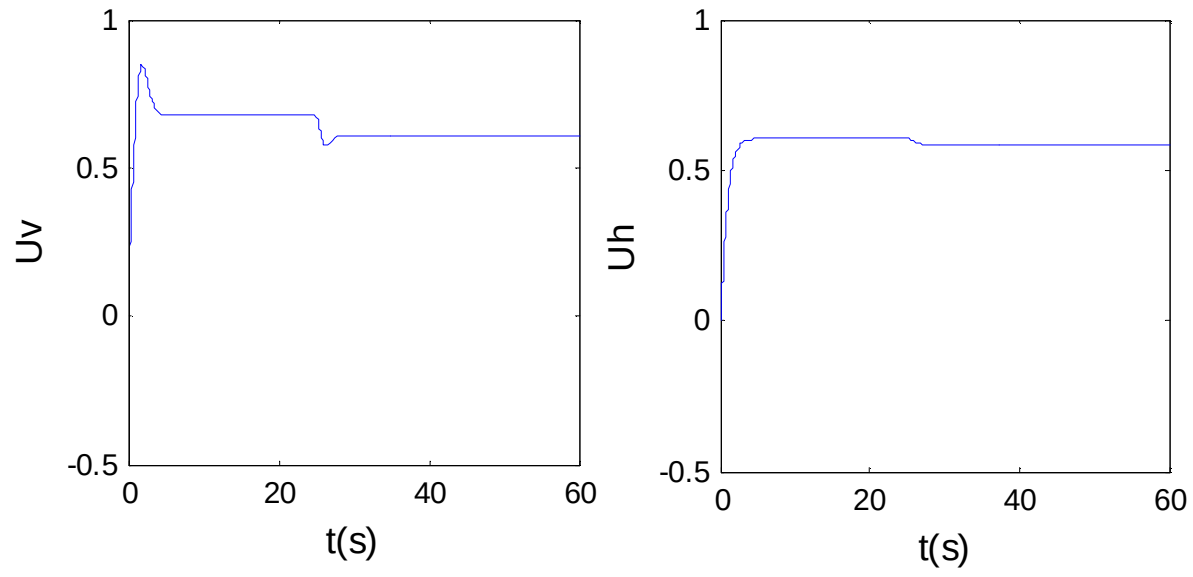

Fig. 10 system commands

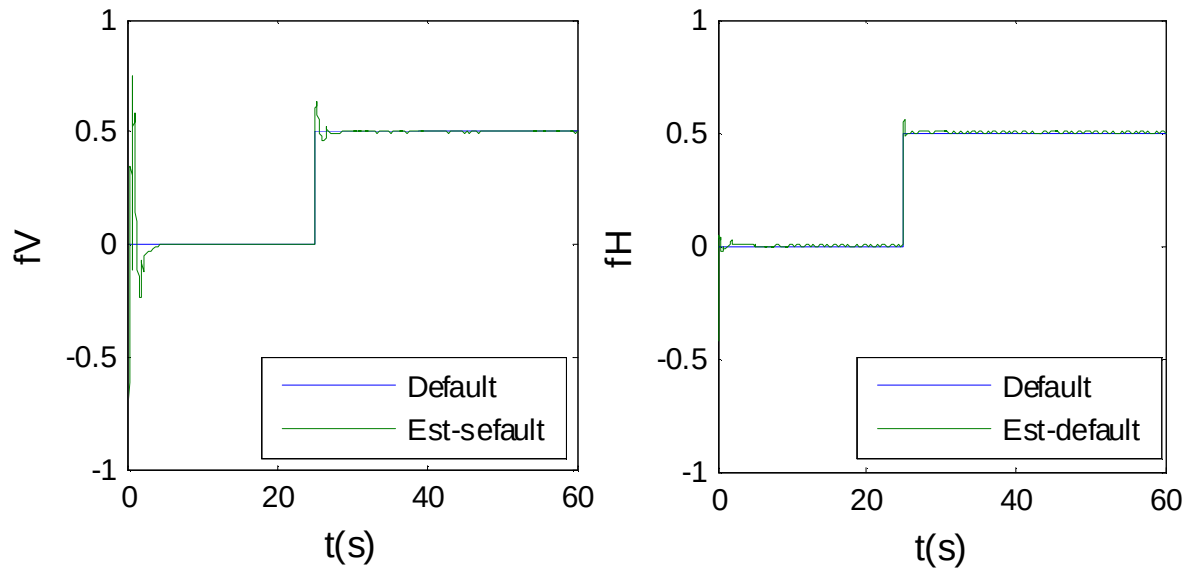

Fig. 11 Defaults: f(t) and the estimated default f̂(t)

The simulation results in the figures (6-11) show that estimates of the state variables are very satisfactory despite the presence of actuator fault which can calculate the FTC control law to reduce the effect of failure on the system. We note that the outputs $\alpha_v$ and $\alpha_h$ correctly follow the variations references.

## V. Conclusions

In this paper, we presented stabilizing control laws synthesis by H∞ and state feedback. Firstly, we start by the development of the dynamic model of the TRMS taking into account the different physics phenomena. After we are interested to propose the FTC controller based on H∞, this controller is designed such that it can stabilize the faulty plant using H∞ theory and LMIs; this method was suitable for partial actuator faults. Simulation results also validate that the presented FTC has a satisfactory tracking performance and is robust to the external disturbances.